\documentclass[twocolumn,showpacs,preprintnumbers,amsmath,amssymb]{revtex4-1}
\usepackage{graphicx}
\usepackage{textcomp}
\usepackage{xcolor}

\begin{document}
\title{Quantum decay in a topological continuum}
  \normalsize
\author{Stefano Longhi}
\address{Dipartimento di Fisica, Politecnico di Milano and Istituto di Fotonica e Nanotecnologie del Consiglio Nazionale delle Ricerche, Piazza L. da Vinci
32, I-20133 Milano, Italy}

%
\bigskip
\begin{abstract}
The quantum mechanical decay of two or more overlapped resonances in a common continuum is largely influenced by Fano interference, leading to important phenomena such as the existence of bound states in the continuum, fractional decay and quiescent dynamics for single particle decay, and signature of particle statistics in the many-body quantum decay. An overlooked yet essential requirement to observe Fano interference is time reversal symmetry of the bath. Here we consider multilevel quantum decay in a bath sustaining unidirectional (chiral) propagating states, such as in quantum Hall or in Floquet topological insulators, and show that the chiral nature of scattering states fully suppresses Fano interference among overlapping resonances. As a result, there are not bound states in the continuum, quantum decay is complete, and there is not any signature of particle statistics in the decay process. Nonetheless, some interesting features are disclosed in the multilevel decay dynamics in a topological bath, such as the appearance of high-order exceptional points, long quiescent dynamics followed by a fast decay, and the possibility to observe damped non-Hermitian Bloch oscillations.
\noindent

\end{abstract}

\maketitle

\section{Introduction}
The decay of metastable states into a continuum is ubiquitous in many areas of physics. Examples include the quantum mechanical decay of particles, nuclei, atoms, and molecules \cite{r1,r2,r3}, spontaneous emission of a photon from excited states of atoms \cite{r4}, tunneling escape from a potential trap \cite{r5}, photon leakage in integrated optical structures \cite{r6,r7}, dark energy decay in cosmological models \cite{r8}, etc.
The decay of a single discrete state is well described, in the weak coupling and markovian approximation \cite{r4}, by an exponential decay law, leading to symmetric (Breit-Wigner) resonance states which are ubiquitous in most of the quantum mechanical decay processes. However, deviations from exponential decay occur at both short and long time scales as a result of memory effects \cite{r2}, which can be exploited to either accelerate or decelerate the quantum mechanical decay via frequent observations \cite{r8bis}. When two or more discrete states decay into a common continuum, interesting phenomena can be observed as a result of Fano interference among overlapping resonances \cite{r9,r10}. These include limited decay and the existence of bound states in the continuum (also referred to as dark states \cite{r3}) for perfect destructive interference of different decay channels \cite{r3,r4,r11,r12,r13,r14}, qualitative changes in long-time decay behavior \cite{r15}, damped oscillations \cite{r16}, resilient periods followed by decay bursts \cite{r17}, exponential-power law decay near an exceptional point (EP) \cite{r18,r18bis}, particle statistics dependence in many-body quantum decay \cite{r19,r20}, and interference in entanglement decay \cite{r21}. A prototypal model of quantum decay is provided  by the Fano-Anderson (or Friedrichs-Lee) model, which describes the coupling of one or more discrete states to a common one-dimensional continuum \cite{r3,r11,r14,r20,r22,r23}.\\
Previous studies have mostly considered quantum decay into a continuum with time reversal symmetry. In effective one-dimensional models, this means that scattering states into which the discrete state decays propagate bidirectionally in the bath. However, in topological matter, such as in quantum Hall systems and topological insulators, scattering states can show unidirectional (chiral) propagation as a result of time reversal symmetry breaking. Chiral edge states in topological insulators have  received a huge interest in past recent years in several areas of physics, ranging from condensed-matter physics to photonics and beyond (see, for instance, \cite{r24,r25,r26} and references therein). So far, quantum decay in a topological continuum and the role of chiral edge states onto Fano interference and decay dynamics remain largely unexplored.\\
In this work we study theoretically quantum decay of discrete states into a common topological continuum, which is effectively modeled by a one-dimensional tight-binding lattice with broken time reversal symmetry sustaining unidirectional propagating scattering states. Such a continuum can describe, for example, a two-dimensional quantum Hall insulator or a Floquet topological insulator when the energy of the discrete states fall in a topological gap of the two-dimensional crystal and the decay arises because of the coupling with the chiral edge states in the topological gap. Owing to the chiral nature of  the scattering states, it is shown that Fano interference among overlapping resonances is fully suppressed and bound states in the continuum can not be found, regardless the number and spectral coupling shape of the discrete states. Nonetheless, some interesting features arising from the decay in the topological continuum are found within the markovian approximation. These include the appearance of high-order exceptional points, i.e. non-Hermitian degeneracies \cite{r27} in the effective non-Hermitian description of the decay dynamics, long quiescent dynamics followed by an abrupt decay, the possibility to observe damped non-Hermitian Bloch oscillations when a gradient field is applied to the discrete states, and the independence of quantum decay on particle statistics in the many-body decay dynamics.

\section{Quantum decay of interfering resonances: Fano-Anderson model and non-Hermitian dynamics}
Quantum mechanical decay of interfering resonances can be studied rather generally by means of the Fano-Anderson (or Friedrichs-Lee) model \cite{r3,r11,r14,r22,r23}, which describes the decay of $N$ discrete states of frequencies $\omega_1$, $\omega_2$, ..., $\omega_N$ coupled to a common one-dimensional continuum of states [Fig.1(a)]. As shown in many previous works, in the weak-coupling and markovian approximations the decay dynamics of the interfering resonances in a broad continuum can be described by an effective non-Hermitian Hamiltonian. Here we briefly review for the sake of completeness the main model and results of the analysis.\\
The second-quantization Hamiltonian of the $N$-level Friedrichs-Lee model (with $\hbar=1$) reads 
\begin{eqnarray}
\hat{H} & = & \sum_{\alpha=1}^N \omega_{\alpha} \hat{c}^{\dag}_{\alpha} \hat{c}_{\alpha}+ \int dk \; \omega(k) \hat{c}^{\dag}(k) \hat{c}(k) \nonumber \\
& + & \sum_{\alpha=1}^N \int dk \left[ g_{\alpha}(k) \hat{c}^{\dag}_{\alpha} \hat{c}(k)+ g_{\alpha}^*(k) \hat{c}^{\dag}(k) \hat{c}_{\alpha} \right] \\
& \equiv & \hat{H}_{bs}+\hat{H}_{bath}+\hat{H}_{int} \nonumber
\end{eqnarray}
where: $\hat{c}_{\alpha}$,  $\hat{c}^{\dag}_{\alpha}$ are the annihilation and creation operators of  particles for the bound states at energies $\omega_{\alpha}$ ($\alpha=1,2,...,N$);  $\hat{c}(k)$, $\hat{c}^{\dag}(k)$ are the annihilation and creation operators of particles in the effective one-dimensional continuum of states at the energy $\omega(k)$, described by a continuum index $k$ (for example the Bloch wave number if the bath is a tight-binding continuum); and $g_{\alpha}(k)$ is the spectral coupling function between the $\alpha$-th discrete level and the continuum. The operators  $\hat{c}_{\alpha}$,  $\hat{c}^{\dag}_{\alpha}$,  $\hat{c}(k)$, $\hat{c}^{\dag}(k)$  satisfy the usual commutation/anti-commutation relations of bosonic/fermionic particles.
Note that the Hamiltonian $\hat{H}$ commutes with the particle number operator $\hat{G}=\sum_{\alpha} \hat{c}_{\alpha}^{\dag} \hat{c}_{\alpha}+ \int dk \hat{c}^{\dag}(k) \hat{c}(k)$, which is thus a constant of motion.
In the single-particle case $G=1$, particle statistics is not of relevance and the state vector of the system can be expanded as 
\begin{equation}
|\psi(t) \rangle=\sum_{\alpha=1}^{N} c_{\alpha}(t) \hat{c}^{\dag}_{\alpha} |0 \rangle+\int dk c(k,t) \hat{c}^{\dag}(k) | 0 \rangle
\end{equation}
where $c_{\alpha}(t)$ and $c(k,t)$ are the probability amplitudes to find the particle at the $\alpha$-th discrete level  or in the continuum, respectively. 
 The evolution of the amplitude probabilities is governed by the coupled equations 
\begin{eqnarray}
i \frac{dc_{\alpha}(t)}{dt} & = &  \omega_{\alpha} c_{\alpha}+ \int dk g_{\alpha}(k) c(k,t) \\
i \frac{\partial {c}(k,t)}{\partial t} & = & \omega(k) c(k,t)+ \sum_{\alpha=1}^N g^*_{\alpha}(k) c_{\alpha}(t).
\end{eqnarray}
Assuming that at initial time the continuum is in the vacuum state, the degrees of freedom of the continuum can be formally eliminated from Eqs.(3) and (4), yielding a set of integro-differential equations for the occupation amplitudes $c_{\alpha}$. In the Weisskopf-Wigner (markovian) approximation, i.e. by considering the weak-coupling limit $g_{\alpha} \rightarrow 0$ and assuming a broad and non-structured continuum into which the frequencies $\omega_{\alpha}$ are embedded, the decay dynamics of interfering resonances is described by the following effective non-Hermitian Schr\"odinger equation in the subspace of discrete states (see, for instance, \cite{r3,r17,r18bis,r29})
\begin{equation}
i \frac{dc_{\alpha}}{dt} \simeq \sum_{\beta=1}^{N} \mathcal{H}_{\alpha,\beta}c_{\beta}(t).
\end{equation} 
The elements of the $N \times N$ non-Hermitian matrix $\mathcal{H}$ are given by
\begin{equation}
\mathcal{H}_{\alpha,\beta}= \omega_\alpha \delta_{\alpha,\beta}-i \Delta_{\alpha,\beta}
\end{equation}
where 
\begin{eqnarray}
\Delta_{\alpha,\beta}  & = & \int_0^{\infty} d \tau \int dk g_{\alpha}(k) g_{\beta}^*(k) \exp\{i[\omega_{\beta}-\omega(k)] \tau \}.
\end{eqnarray}
The eigenvalues and corresponding eigenvectors of  $\mathcal{H}$ fully  capture the decay dynamics and Fano interference among overlapped resonances (when $\omega_{\alpha}$ are closely spaced). In particular, the interference among the various decay channels can accelerate or decelerate the decay dynamics, and for perfect destructive interference the real part of one or more eigenvalues of $\mathcal{H}$  can vanish, signaling the existence of bound states in the continuum (dark states) and fractional (limited) decay.

\section{Quantum decay and Fano interference in a topological continuum}
 In the most common cases, the Hamiltonian $\hat{H}_{bath}$ describing the continuum of states satisfies time-reversal symmetry and sustains bidirectional propagating (scattering) states. For example, if the bath is a tight-binding one-dimensional crystal, i.e. a quantum wire [Fig.1(b)], $k$ is the Bloch wave number $(-\pi \leq k < \pi$) and the dispersion relation $\omega(k)$ of the tight-binding lattice band satisfies the condition 
\begin{equation}
\omega(-k)=\omega(k)
\end{equation}
because of time-reversal symmetry [Fig.1(c)]. This implies $(d \omega / dk)(-k)=-(d \omega / dk)(k)$, i.e. the continuum sustains bidirectional propagating states at each frequency in the allowed band. The simplest situation, which has bee studied in details in Ref.\cite{r29}, is that of a continuum consisting of a tight-binding one-dimensional lattice with nearest-neighbor hopping amplitude $\kappa$, corresponding to the dispersion relation
\begin{equation}
\omega(k)=2 \kappa \cos (k)
\end{equation}
 and $N$ discrete levels side-coupled to the lattice at sites $n_1,n_2,..,n_N$ with coupling constants $\kappa_1$, $\kappa_2$,..., $\kappa_N$ [see Fig.1(b) and (c)]. Here $\kappa$ is the hopping amplitude between nearest-neighbor sites in the quantum wire. Owing to time reversal symmetry and the bidirectional nature of transport in the lattice, two discrete levels with nearly-degenerate resonance frequency are indirectly coupled via the bidirectional scattered states in the continuum, resulting in a Fano-type interference.\\
In this work we consider the case where the continuum corresponds to {\it unidirectional} propagating states solely, i.e. for which the group velocity $(d \omega/dk)$ has a defined sign (either positive or negative) within the entire Brillouin zone. This case is typically found whenever the bath is a topological continuum, such as a quantum Hall or a Floquet topological insulator with some edges (see Sec.IV below), and the frequencies $\omega_{\alpha}$ of the discrete states are embedded in a topological gap of the insulator: the discrete levels are coupled to the chiral edge states of the topological insulator (and not to the bulk states), which thus acts as an effectively one-dimensional bath (a quantum wire) sustaining unidirectional propagating states solely. The decay of the discrete states coupled to the chiral edge states of the topological insulator can be effectively modeled by considering in Eq.(1) a one-dimensional tight-binding metacrystal with long-range hopping and broken time-reversal symmetry \cite{r30}, which realizes unidirectional transport in the quantum wire. For the sake of definiteness, we consider a rather simple model, schematically shown in Fig.1(b), where $N$ discrete sites are attached to a one-dimensional metacrystal with long-range hopping. In the Wannier basis representation, the Hamiltonian of the system reads
\begin{eqnarray}
\hat{H} &= & \sum_{\alpha=1}^N \omega_{\alpha} \hat{c}^{\dag}_{\alpha} \hat{c}_{\alpha}+ \sum_{l,n} (\theta_{n-l} \hat{A}^{\dag}_n \hat{A}_l+H.c.) \nonumber \\
& + & \sum_{\alpha=1}^{N} (\kappa_\alpha \hat{c}^{\dag}_{\alpha} \hat{A}_{n_{\alpha}}+H.c.)
\end{eqnarray}
where $\hat{A}^{\dag}_n$ is the particle (bosonic/fermionic) creation operator at the $n-th$ Wannier site of the tight-binding lattice, $\theta_{n-l}$ is the hopping rate between sites $n$ and $l$ in the lattice, and $\kappa_\alpha$ is the hopping rate of the $\alpha$-th discrete site $|\alpha \rangle$ attached to the $n_{\alpha}$ site of the lattice [Fig.1(b)]. The dispersion curve of the tight-binding lattice band is given by 
\begin{equation}
\omega(k)= \sum_{n=-\infty}^{\infty} \theta_{n} \exp(-ink), 
\end{equation}
which is assumed to satisfy the constraint $(d \omega / d k)>0$ to ensure unidirectional propagation of scattering states [Fig.1(d)]. The Fano-Anderson Hamiltonian (1) is readily obtained by writing Eq.(10) in the Bloch basis (rather than in the Wannier basis). By letting 
\begin{equation}
\hat{c}(k)=\frac{1}{\sqrt{2 \pi}} \sum_{-\infty}^{\infty} \hat{A}_n \exp(-i k n)
\end{equation}
where $-\pi \leq k < \pi$ is the Bloch wave number, the Hamiltonian (10) is readily transformed into Eq.(1) with $\omega(k)$ given by Eq.(11) and with the following spectral coupling functions $g_\alpha(k)$ 
\begin{equation}
g_{\alpha}(k)=\frac{\kappa_{\alpha}}{\sqrt{2 \pi}} \exp(ik n_{\alpha}).
\end{equation}
The elements of the non-Hermitian matrix (6) can be computed using Eqs.(7) and (13). The calculation, which is detailed in the Appendix A, yields
\begin{equation}
\mathcal{H}_{\alpha, \beta} \simeq \left\{
\begin{array}{cc}
 \omega_{\beta}-i\kappa_{\beta}^{2} /(2 v_{\beta}) & \alpha=\beta \\
 -i ( \kappa_{\alpha} \kappa_{\beta} / v_{\beta}) \exp[i k_ \beta(n_{\alpha}-n_{\beta}]  & n_{\alpha} > n_{\beta} \\
 0 & n_{\alpha} < n_{\beta}
\end{array}
\right.
\end{equation} 
where $k_{\beta}$ and $v_{\beta}$ are defined by [see Fig.1(d)]
\begin{equation}
\omega(k_{\beta})=\omega_{\beta} \; , \; \; v_{\beta}=(d \omega / d k)_{k_{\beta}}>0.
\end{equation}

 \begin{figure}
\includegraphics[scale=0.6]{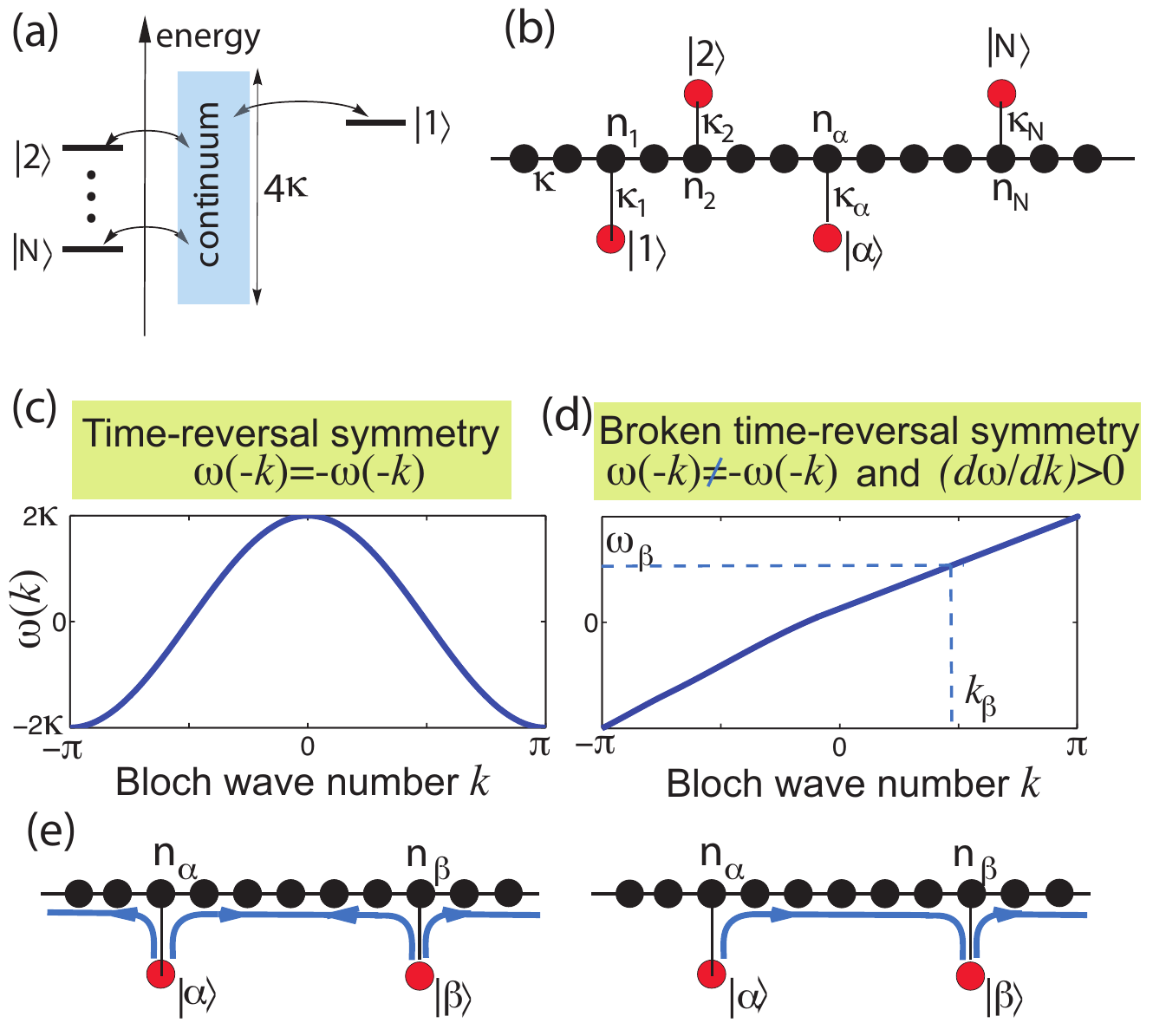}
\caption{(Color online) (a) Schematic of the multilevel Fano-Anderson model. (b) A tight-binding lattice (a quantum wire) with side-coupled $N$ discrete sites realizes quantum mechanical decay of $N$ interfering resonances in a tight-binding continuum. The frequencies of the discrete states are fully embedded in the broad tight-binding lattice band. (c,d) Dispersion relation $\omega(k)$ of the continuum (Bloch) states in (c) a one-dimensional lattice with time-reversal symmetry [$\omega(k)=2 \kappa \cos k$ in the nearest-neighbor approximation], and (d) in a topological continuum with chiral edge states. In (c) destructive Fano interference and bound states in the continuum can arise because of bidirectional propagation in the continuum, whereas in (d) coupling is unidirectional and bound states in the continuum are prevented [see panel (e)].}
\end{figure}

It is worth comparing the expression of the effective non-Hermitian matrix $\mathcal{H}$ given by Eq.(14) with the one corresponding to a tight-binding bidirectional continuum with the usual dispersion relations $\omega(k)=2 \kappa \cos (k)$ [Eq.(9)] satisfying time-reversal symmetry. The elements of the non-Hermitian matrix $\mathcal{H}$ for the bidirectional quantum wire have been calculated in Ref.\cite{r29}, and are given in Appendix B for the sake of completeness.\\
Note that, for the unidirectional continuum with broken time reversal symmetry, $\mathcal{H}$ is  a lower triangular matrix, i.e. the elements $\mathcal{H}_{\alpha,\beta}$ in the upper diagonals $\beta>\alpha$ vanish [see Eq.(14)], indicating that the eigenvalues of $\mathcal{H}$ are the elements on the main diagonal. Such a striking result follows from the unidirectional (chiral) transport in the bath, yielding  {\em unidirectional} coupling between any of two discrete states $|\alpha \rangle$ and $|\beta \rangle$: contrary to a bath sustaining bidirectional transport, in case of a chiral bath excitation that decays from a site $| \alpha \rangle$ into the continuum is detected by any other site $|\beta \rangle$ in the chain if and only if $n_{\beta} > n_{\alpha}$ [see Fig.1(e)]. Such a behavior has a strong impact into the decay dynamics as compared to a usual bidirectional continuum, with some remarkable results that are discussed below. \par
(i) {\it Absence of bound states in the continuum and complete decay.} The unidirectional nature of coupling among the discrete levels inhibits destructive interference among different decay channels (i.e. Fano interference), resulting in the complete decay of any initial excitation into the continuum and the absence of bound states in the continuum. In fact, a bound state in the continuum would correspond to an eigenvalue of $\mathcal{H}$ with vanishing imaginary part. While this is possible for a bidirectional continuum under certain conditions, corresponding to complete destructive quantum interference of decay channels (see e.g. \cite{r29} and Appendix B), in the unidirectional continuum from Eq.(14) it follows that the imaginary part of any eigenvalue $\mathcal{H}_{\alpha, \alpha}$ of $\mathcal{H}$ is strictly negative, resulting in a complete decay. Note also that, owing to the unidirectional nature of the coupling, the decay of level $|1 \rangle$ is always described (within the markovian approximation) by an exponential law and it is not influenced by the presence of the other discrete levels.   
\par
(ii) {\it High-order exceptional points.} A particular regime is attained whenever $\omega_{\alpha}$ and $\kappa_{\alpha}$ are independent of index $\alpha$, i.e. all discrete states have the same resonance frequency and are coupled with the same strength to the continuum. In this case there is a non-Hermitian coalescence of all eigenvalues and corresponding eigenvectors of the non-Hermitian matrix $\mathcal{H}$, i.e. an exceptional point of order $N$ \cite{r27} arises in the multilevel decay dynamics. This is a rather interesting result, since in most common multilevel decay into a non-topological continuum high-order exceptional points are rarely  found and under very special conditions. The presence of high-order exceptional points results in a high non-normal behavior of the decay dynamics and a characteristic exponential power-law decay, discussed in Ref.\cite{r18bis}.\\ 
(iii) {\it Quiescent dynamics followed by an abrupt decay in the topological continuum.} The absence of bound states in the continuum makes the quantum decay into the topological bath complete. However, an interesting feature is that, for a large number $N$ of discrete states, the system can be prepared at initial time in a coherent superposition of the discrete states such that the dynamics is quiescent for a certain time interval $\tau$, where decay is negligible, after which an abrupt decay into the continuum is observed. The quiescence time $\tau$ increases almost linearly with the number $N$ of discrete levels. 
To illustrate such a behavior, let us introduce the non-decaying (survival) probability at time $t$, defined as $P(t)=\sum_{\alpha=1}^N |c_{\alpha}(t)|^2$, with $P(0)=1$. For a given initial condition $\mathbf{c}(0) \equiv (c_1(0),c_2(0),...,c_N(0))^T$,  the upper bound for $P(t)$ is given by $P(t) \leq \sigma_{\max}(t)$,
where $\sigma_{\max}(t)$ is the largest eigenvalue of the matrix $\mathcal{A}^{\dag} (t) \mathcal{A}(t)$ and where we have set $\mathcal{A} (t) \equiv \exp(-i \mathcal{H} t)$ \cite{r18bis}. {For a fixed time $t>0$, the largest  value $\sigma_{\max}(t)$  is assumed for the initial excitation $\mathbf{c}(0)$ of the system which is the eigenvector of $\mathcal{A}^{\dag}(t) \mathcal{A}(t)$ with eigenvalue $\sigma_{\max}(t)$.
 Quiescent dynamics followed by a fast decay occurs when $\sigma_{\max}(t)$ versus $t$} remains very close to one for some interval $0<t<\tau$, after which the decay starts with an abrupt drop of the survival probability. {The resilience time $\tau$ can be thus defined as the largest time $t=\tau$ such that the upper bound $\sigma_{\rm max}(t)$ of the survival probability remains larger than a reference value $P_b$ in the entire time interval $(0, \tau)$, i.e. $\sigma_{\max}(t) > P_b$ for $0<t< \tau$ and $\sigma_{\max}(t)<P_b$ for $t> \tau$. Clearly, the choice of the reference value $P_b$ is  not unique and can be empirically set looking at the actual profile of the decay law, which can be more or less steep. This means that there is some uncertainty in the definition of $\tau$.}
 Figure 2(a) shows a typical behavior of the upper bound $\sigma_{\max}(t)$ of the survival probability $P(t)$ versus $t$ corresponding to multilevel quantum decay of $N=20$ discrete levels in a topological continuum (solid curve) and in an ordinary tight-binding continuum (dashed curve). Clearly, for the case of a topological continuum a quiescent dynamics of the survival probability for a time interval $\tau$ is observed, followed by a rapid decay into the bath. {The numerically-computed behavior of the quiescence time $\tau$ versus $N$ for three values of the reference level $P_b$ is shown in Fig.2(b). The figure clearly shows that the resilience interval increases almost linearly with the number of discrete levels, a result that does not sensitively depend on the precise value of the reference level $P_b$}. To observe resilience dynamics in the decay process, the system should be prepared at initial time $t=0$ is a coherent superposition of amplitudes $c_{\alpha}(0)$ given by the eigenvector of $\mathcal{A}^{\dag} \mathcal{A}$ for some fixed value of $t \leq \tau$. An example of quiescent dynamics followed by an abrupt decay into the bath, for a number $N=20$ of discrete levels, is shown in Fig.2(c). {Here the system is initially prepared in a coherent state, shown in the inset of Fig.2(c), corresponding to the eigenstate of the matrix $\mathcal{A}^{\dag}(\tau) \mathcal{A}(\tau)$ with $ \tau \simeq 28$ (the resilience time for $N=20$). Note that such a state corresponds to the excitation of a few discrete sites around the site $\alpha=3$, near the left edge of the impurity chain. To physically explain the absence of the decay into the continuum until the time $t \sim \tau$, is is worth looking at the temporal evolution of the amplitudes $|c_{\alpha}(t)|$ in the various discrete sites, which is depicted in Fig.2(d). The figure clearly shows that the excitation remains trapped in the chain of discrete sites and shifts in time at a constant speed until the right-edge discrete site is reached, after which an abrupt decay into the continuum is observed. Such a simple physical picture explains the resilience against quantum decay despite the absence of bound states and why the resilience time $\tau$ increases linearly with the number $N$ of discrete sites.}
  \par
   \begin{figure}
\includegraphics[scale=0.28]{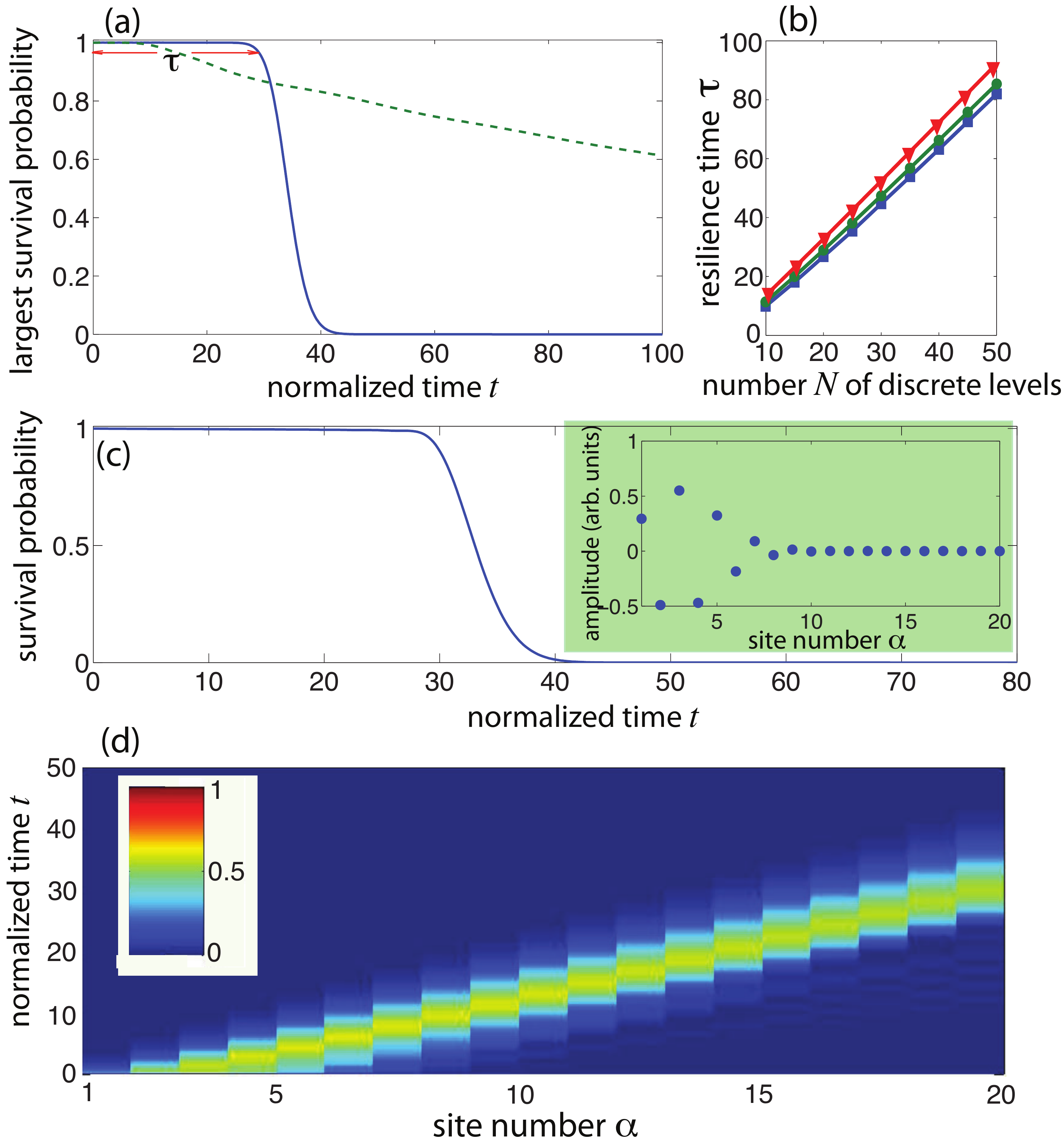}
\caption{(Color online) Quiescent dynamics in multilevel quantum decay. (a) Behavior of the upper bound $\sigma_{\max}(t)$ of the survival probability $P(t)$ versus time $t$ in the system of Fig.1(b) comprising $N=20$ degenerate discrete sites side-coupled to a tight-binding continuum with $\omega_{\alpha}=0$, $n_{\alpha}=\alpha$ and $\kappa_{\alpha}$ independent of $\alpha$. The solid curve refers to a topological continuum with dispersion curve as in Fig.1(d),  whereas the dashed curve corresponds to a tight-binding continuum with dispersion relation as in Fig.1(c). Time is normalized to $\kappa_{\alpha}^2 / (2 v_{\alpha})$ in the former case, whereas it is normalized to $\kappa_{\alpha}^2 / (2 \kappa)$ in the latter case. Clearly, the survival probability in the quantum decay in the topological continuum  shows a quiescent dynamics for a time interval $\tau$, after which an abrupt decay is observed. The resilience time $\tau$, defined as the time $\tau$ such that $\sigma_{\max}(\tau)=P_b$, turns out to be an almost linearly increasing function of the number $N$ of the discrete levels, as shown in panel (b) for three different values of the reference level $P_b$ (squares: $P_b=0.996$; circles: $P_b=0.97$; triangles: $P_b=0.5$). (c) Example of the survival probability for $N=20$ levels decaying in the topological continuum for the coherent initial preparation of the system in the state shown in the inset (amplitudes $c_{\alpha}$ at initial time $t=0$). The time evolution of the amplitudes $|c_{\alpha}(t)|$ at various discrete sites is depicted in panel (d). Note that the excitation remains trapped in the impurity sites and shifts in time until the right edge site $\alpha=20$ is reached, after which an abrupt decay into the quantum wire occurs.}
\end{figure}
(iv) {\it Non-Hermitian damped Bloch oscillations.} An interesting dynamical behavior is observed when a gradient field $C$ is applied to the $N$ discrete levels, i.e. the resonances are equally-spaced in frequency $\omega_{\alpha}=\omega_1+C (\alpha-1)$, and they are coupled to the topological continuum with the same coupling strength and group velocity, i.e. $\kappa_{\alpha}$ and $v_{\alpha}$ are independent of $\alpha$. In this case the eigenvalues of the non-Hermitian matrix $\mathcal{H}$ describing the decay dynamics read
\begin{equation}
\lambda_{\alpha}=\omega_{1}+C (\alpha-1)-i \Delta
\end{equation}
($\alpha=1,2,...,N$) where $\Delta \equiv \kappa_{\alpha}^2/(2 v_{\alpha})$ is the common decay rate. Since the real part of the eigenvalues are equally spaced by $C$, the temporal dynamics is periodic with period $T_B= 2 \pi/C$, enveloped by an exponential decay with decay rate $\Delta$. Such a dynamical behavior corresponds to damped Bloch oscillations, which are peculiar to non-Hermitian lattices with unidirectional hopping \cite{r31}. Note that, contrary to usual Bloch oscillations in Hermitian lattices, here edge effects do not smear out the periodicity of the oscillations, which are observed even when dealing with few discrete levels \cite{r31}. An example of damped Bloch oscillations of $N=6$ discrete levels decaying into a topological continuum is shown in Fig.3. The damped oscillatory behavior with period $T_B$ is clearly observed when detecting the decay behavior of the amplitudes in the discrete levels $\alpha \geq 2$ [Fig.3(a)]. Interestingly, the signature of the damped Bloch oscillations is also visible in the decay behavior of the survival probability $P(t)$ [Fig.3(b)], which is slowed down after each Bloch oscillation period.

\begin{figure}
\includegraphics[scale=0.28]{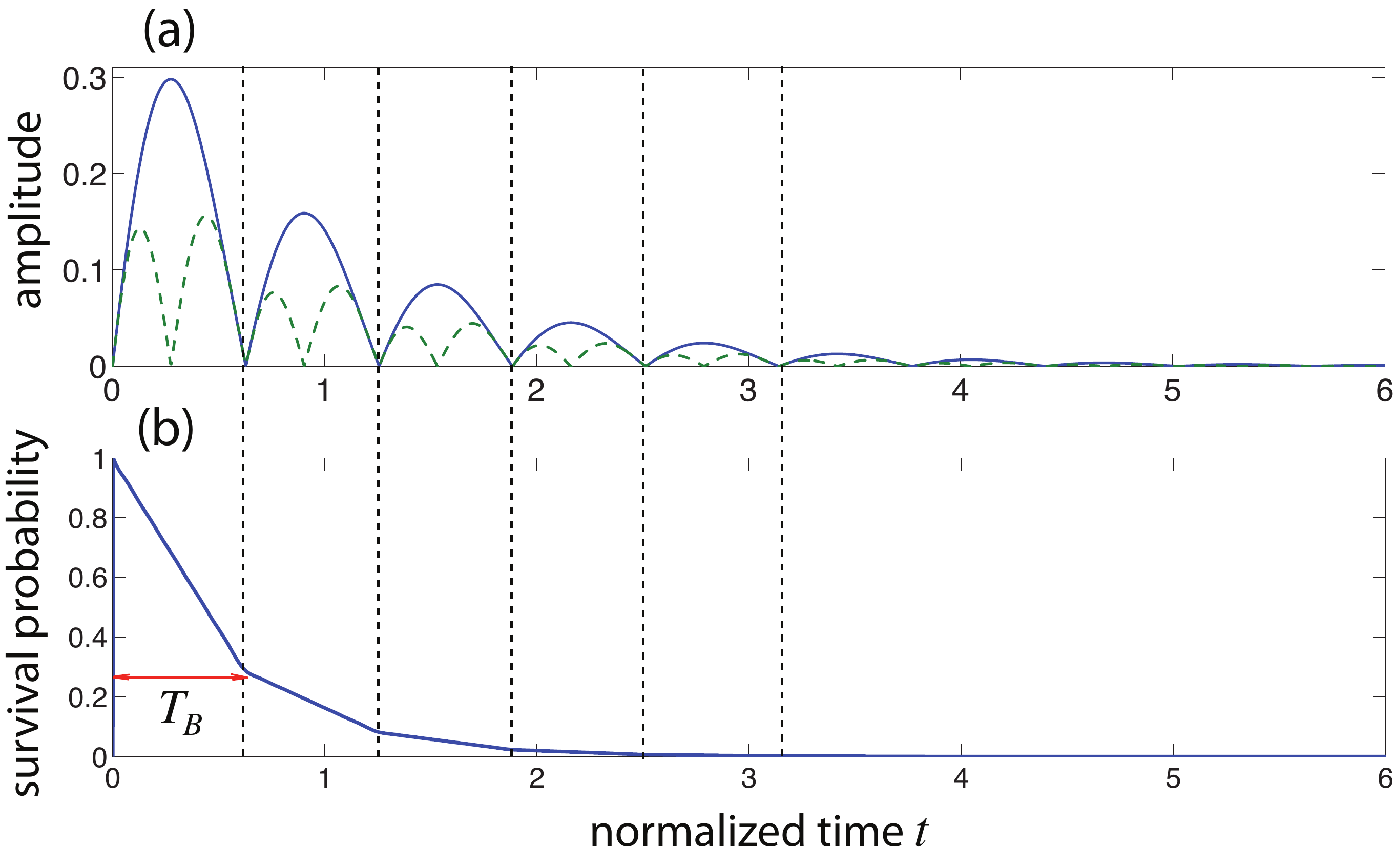}
\caption{(Color online) Non-Hermitian damped Bloch oscillations in multilevel quantum decay in a topological continuum. $N=6$ discrete levels are side coupled to the topological tight-binding continuum with equally-spaced resonance frequencies $C=10 \Delta$, where $\Delta= \kappa_{\alpha}^2 / 2 v_{\alpha}$ is the decay rate. At initial time $t=0$ the system is prepared with the excitation in site $\alpha=1$, i.e. $c_{\alpha}(0)=\delta_{\alpha,1}$. Panel (a) shows the temporal dynamics of the amplitudes $|c_{\alpha}(t)|$ for $\alpha=2$ (solid curve) and $\alpha=3$ (dashed curve), clearly showing a damped oscillatory dynamics with Bloch oscillation period $T_B= 2 \pi /C$. Time is normalized to the inverse of decay rate $1/ \Delta$. The behavior of the survival probability is shown in panel (b).}
\end{figure}

\section{Examples of multilevel decay in topological baths}
Chiral edge states in two-dimensional topological insulators provide a major platform to realize an effective one-dimensional topological continuum showing unidirectional propagating modes. Provided that the energies of the discrete levels are entirely embedded in a topological gap of the crystal, decay arises because of the coupling of the discrete levels  with  the chiral edge states, which thus realize an effective one-dimensional continuum with a dispersion curve as the one shown in Fig.1(d). Chiral edge states emerge in a wide variety of topological quantum and classical systems, such as in quantum Hall systems, in the Haldane model, in Floquet topological insulators and in anomalous Floquet topological insulators to mention a few (see for example \cite{r25,r26,r32} and references therein). Here we discuss multilevel decay in two examples of topological baths, namely in a tight-binding quantum Hall system (the Harper-Hofstadter model), and in anomalous Floquet topological insulators, and compare exact numerical results with the non-Hermitian effective description presented in the previous section. 
\subsection{Quantum decay in the Harper-Hofstadter topological bath}
The celebrated Harper-Hofstadter model \cite{r33} describes charged particles moving in a two-dimensional tight-binding square lattice under a uniform magnetic 
flux per unit cell [Fig.4(a)]. This model realizes a quantum Hall topological insulator where the magnetic flux breaks time reversal symmetry and topological features are characterized by
the  first Chern numbers.  The Harper-Hofstadter model  has been realized in different kinds of synthetic matter with artificial gauge fields, including ultracold atoms in optical lattices \cite{r34}, microwaves circuits \cite{r35}, and 
photons in coupled microring resonators \cite{r36}. In the infinitely-extended lattice, the single-particle energy spectrum depends sensitively
on the magnetic flux per unit cell and the tight-binding band of the square lattice splits into narrow magnetic
bands. At high magnetic fields, a self-similar (fractal) energy spectrum does emerge,
which is known as the Hofstadter butterfly. The Hamiltonian of the Harper-Hofstadter model reads
\begin{equation}
\hat{H}_H= \kappa \sum_{n,m} \left\{ \hat{a}^{\dag}_{n+1,m} \hat{a}_{n,m}+ \exp(i \varphi n) \hat{a}^{\dag}_{n,m+1} \hat{a}_{n,m}+H.c. \right\}
\end{equation}
where $\kappa$ is hopping rate between adjacent sites in the square lattice and $\varphi$ is the magnetic flux in each plaquette. 
Here we consider the case of a magnetic flux $\varphi=\pi/2$, corresponding to band splitting into four magnetic bands with two wide topological gaps, shown in  Fig.4(c). 
 When the infinite two-dimensional lattice is truncated at one edge in the $n$ direction, as in Fig.4(b), chiral edge states in each one of the two wide topological gaps arise with a dispersion curve $\omega(k)$ which can be computed from the eigenvalue Harper equation 
\begin{equation} 
 \kappa(A_{n+1}+A_{n-1})+ 2 \kappa \cos (k+ n \varphi) A_n=\omega(k) A_n
\end{equation}
with the boundary conditions $A_0=0$ and $A_{n} \rightarrow 0$ as $n \rightarrow + \infty$.
The chirality of the edge states is opposite in the two gaps, as shown in Fig.4(c).  
  \begin{figure}
\includegraphics[scale=0.28]{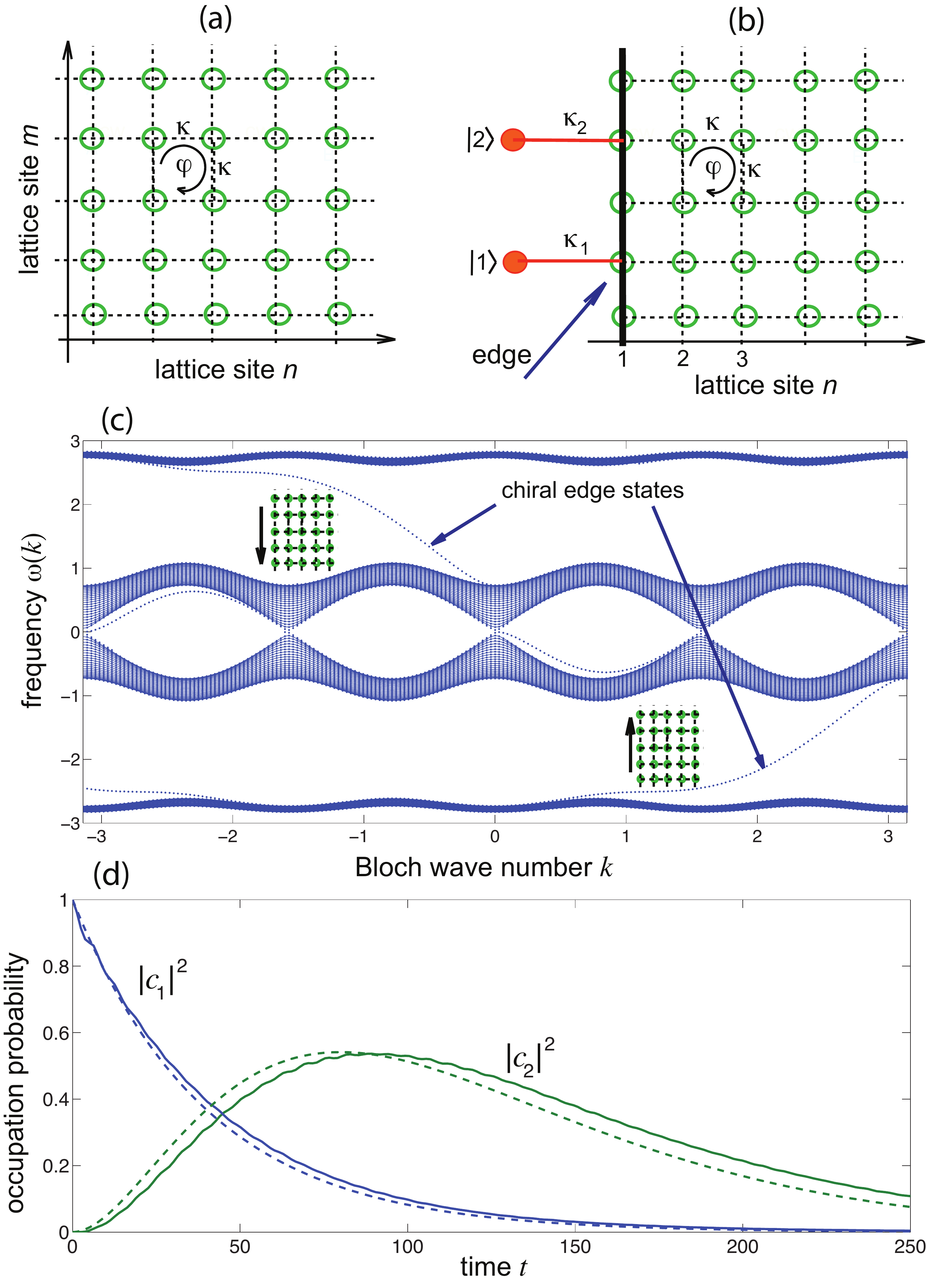}
\caption{(Color online) Quantum mechanical decay in a quantum Hall topological bath. (a) Infinitely-extended tight-binding square lattice in the presence of a magnetic flux $\varphi$ per lattice unit cell that breaks time reversal symmetry. $\kappa$ is the nearest-neighbor hopping amplitude. (b) Semi-infinite square lattice with an edge at $n=1$. (c) Energy spectrum of the semi-infinite square lattice for a magnetic flux $\varphi= \pi/2$ and for $\kappa=1$. Note that there are four magnetic bands with two wide topological gaps. Chiral edge states, with opposite group velocities, are found with energies in the upper and lower wide gaps. If the frequencies $\omega_{\alpha}$ of the discrete levels, side-coupled to the edge of the square lattice, are entirely embedded in the lower wide magnetic gap, the chiral edge states with positive group velocities act as an effective one-dimensional topological bath sustaining unidirectional propagating modes. (d) Decay dynamics (behavior of the occupation probabilities $|c_{1}(t)|^2$ and $|c_{2}(t)|^2$ versus time $t$) for $N=2$ discrete levels with frequencies $\omega_{1}=\omega_2=-1.5$ side-coupled, with equal coupling constants $\kappa_1=\kappa_2=0.2$, to the sites $(n,m)=(1,0)$ and $(n,m)=(1,3)$ of the square lattice. At initial time the system is prepared in the discrete state $|1 \rangle$. Solid and dashed curves refer to the exact numerical results and to the approximate non-Hermitian description, respectively. The values of $k_{\beta}$ and of the group velocity $v_{\beta}$ used in the non-Hermitian matrix $\mathcal{H}$ are obtained from the dispersion curve $\omega(k)$ of the chiral edge states and the condition $\omega(k_{\beta})=\omega_{1,2}$, and read $k_{\beta} \simeq 2.536$ and $v_{\beta}=(d \omega / dk)_{\omega_{1,2}} \simeq 1.6$. Note that, owing to unidirectional Fano interference, the decay of level $|1 \rangle$ is almost exponential and it is not influenced by the presence of the discrete level $|2 \rangle$.}
\end{figure}
When a number  $N$ of discrete levels $|1 \rangle$, $|2 \rangle$, ... are side-coupled to the edge of the lattice [$N=2$ in Fig.4(b)] and their frequencies are embedded in either one of the two wide topological gaps, the semi-infinite two-dimensional square lattice, sustaining unidirectional edge states in either one of the two gaps, acts as an effective one-dimensional topological bath with a dispersion curve $\omega(k)$ like in Fig.1(d), into which the discrete levels can decay. A positive group velocity $(d \omega /d k)>0$ is obtained when the frequencies of the discrete levels are embedded in the lower topological gap. As an example, Fig.4(d) shows the numerically-computed decay dynamics of two discrete levels, attached at the edge at sites $(n,m)=(1,0)$ and $(n,m)=(1,3)$, into the chiral edge states in the lower topological gap. Parameter vaues used in the simulations are $\kappa=1$ (hopping rate in the square lattice), $\kappa_1=\kappa_2=0.2$ (coupling constants of discrete levels with the lattice), and $\omega_1=\omega_2=-1.5$ (resonance frequencies of the discrete levels). Solid and dashed curves in the figure depict the exact numerical results (solid curves) and the approximate ones (dashed curves) obtained by the effective non-Hermitian model [Eqs.(5,6,14) in previous sections], respectively. Note that, owing to the unidirectional nature of the coupling between the discrete levels, the decay of level $|1 \rangle$ is almost exponential and it is not influenced by the presence of the discrete level $|2 \rangle$.

\subsection{Quantum decay in an anomalous Floquet topological bath}
A bath showing unidirectional transport can be realized in Floquet topological insulators, where time reversal symmetry is broken by periodic temporal modulation of the underlying Hamiltonian of a crystal \cite{r32}. An interesting case is the one of anomalous topological insulators \cite{r37}, whose topological classification goes beyond that of static systems. Anomalous Floquet topological insulators have been recently demonstrated in synthetic matter based on photonic \cite{r38} and sound \cite{r39} transport in two-dimensional lattices.  Interestingly, to achieve unidirectional (rectified) transport based on Floquet driving it is enough to consider a quasi-one dimensional lattice \cite{r40}, namely a binary lattice with controlled coupling between nearest-neighbor sites, as demonstrated in an earlier experiment Ref.\cite{r41}. A scheme of multilevel decay in the anomalous Floquet topological bath, consisting of two slowly-driven sublattices A and B into which the side-coupled discrete levels can decay, is shown in Fig.5(a). The modulation cycle, of period $T=2T_1+T_2$, comprises threes steps. In the first step (time interval $0<t<T_1$), alternating dimers of sublattices A and B are coupled by a coupling constant $\kappa$ such that $\kappa T_1=\pi/2$ [solid bonds in Fig.5(a)], while in the second step (time interval $T_1<t<2T_1$) the other dimers are coupled with the same coupling constant. In such two steps the discrete levels are decoupled from the sublattices. Note that, owing to the condition $\kappa T_1=\pi/2$, after the two steps any initial excitation in sublattice A is shifted forward by one site in A with a $\pi$ phase shift, whereas any initial excitation in sublattice B is shifted backward by one site in B with $\pi$ phase shift \cite{r40,r41}. In momentum space, the one-site forward/backward spatial shift is simply described by the operator $\exp(\pm ik+i\pi)$. In the third step of the modulation cycle (time $2T_1< t < T$), the sites in sublattices A and B are decoupled, while the discrete levels $|1 \rangle$, $|2 \rangle$,... are side-coupled to the sites $n_1$, $n_2$, ... of sublattice A by the coupling constants $\rho_1$, $\rho_2$, ... Clearly, assuming that at initial time $t=0$ the bath is in the vacuum state, after each modulation cycle, i.e. at times $t=T,2T,3T,...$ the sublattice B remains in the vacuum state, while excitation can spread in sublattice A, which acts as an effective topological bath sustaining unidirectional propagating Bloch modes. The propagator $\hat{U}$ of the system over one oscillation cycle can be factorized as
   \begin{figure}
\includegraphics[scale=0.28]{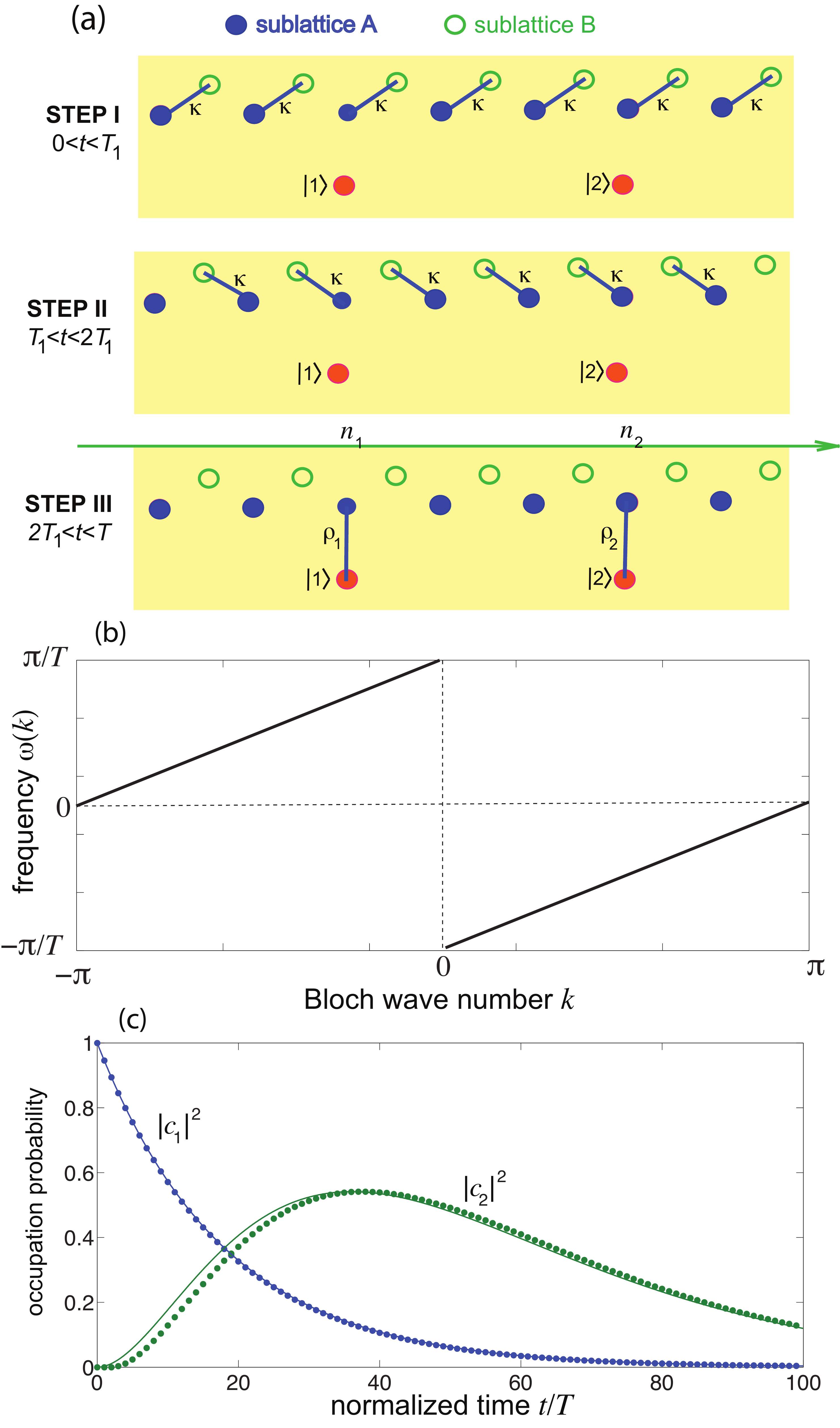}
\caption{(Color online) Multilevel decay in an anomalous Floquet topological insulators. (a) Scheme of the slowly-driven system. The topological bath, comprising two sublattices A and B, is side-coupled to a number $N$ of discrete levels $|1 \rangle$, $|2 \rangle$, ... ($N=2$ in the figure) via sublattice A. The modulation cycle of period $T=2T_1+T_2$ comprises three steps. In the first two steps, each of duration $T_1$, the discrete levels are uncoupled to the bath and alternating dimers of the sublattices are coupled with a hopping amplitude $\kappa$ satisfying the condition $\kappa T_1= \pi/2$. In the third step (time duration $T_2$), the sites in sublattices A and B are decoupled, while the discrete levels $|1 \rangle$, $| 2 \rangle$, ... are coupled to the sites $n_1$, $n_2$, ... of sublattice A with coupling constants $\rho_1$, $\rho_2$, ... (b) Dispersion relation (quasi energy) of the effective topological bath. Note that the group velocity $v=(d \omega / dk)=1/T$ is independent of frequency. (c) Example of two-level decay in the anomalous Floquet topological continuum. Parameter values are $\omega_1=\omega_2=0$, $T_1=T_2=T/3$, $\kappa=3\pi/(2T)$, $\rho_1=\rho_2=0.15 \kappa$, $n_1=0$ and $n_2=2$. At initial time the system is prepared with the excitation in level $|1 \rangle$. The evolution of the occupation probabilities of the two discrete levels at stroboscopic times $t=0,T,2T,...$ as obtained by numerical computation of the exact Floquet dynamics, is shown by bold circles, whereas the approximate continuous-time dynamics governed by the corresponding non-Hermitian Hamiltonian Eq.(14) is depicted by thin solid curves. Note that, owing to unidirectional Fano interference, the decay of level $|1 \rangle$ is almost exponential and it is not influenced by the presence of the discrete level $|2 \rangle$.}
\end{figure}
\begin{equation}
\hat{U}= \hat{U}_{2} \hat{U}_{1}  
\end{equation}
where $\hat{U}_{1}$ describes  the system evolution in the first and second steps of Fig.5(a), i.e. from $t=0$ to $t=2T_1$, whereas $\hat{U}_{2}$ describes  the system evolution in the third step, i.e. from $t=2T_1$ to $t=T$. Since in the first two steps the binary lattice and discrete levels are decoupled, one has
\begin{equation}
\hat{U}_1=\exp (-i \hat{H}_{bath}T- i \hat{H}_{bs} \tau) \equiv \exp(-i \hat{H}_1 T)
\end{equation}
where we have set $\tau \equiv 2T_1$,
\begin{equation}
\hat{H}_{bs} \equiv \sum_{\alpha=1}^N \omega_{\alpha} \hat{c}^{\dag}_{\alpha} \hat{c}_{\alpha} 
\end{equation}
 is the Hamiltonian of the discrete levels and 
\begin{equation}
\hat{H}_{bath} \equiv \int_{-\pi}^{\pi} dk \omega(k)  \hat{c}^{\dag}(k)  \hat{c}(k)
\end{equation}
is the effective Hamiltonian of the bath.
In the above equations, $\omega_{\alpha}$ is the resonance frequency of the discrete level $| \alpha \rangle$ and $\hat{c}^{\dag}_{\alpha}$ the corresponding creation operator of the excitation in the level, 
$\hat{c}^{\dag}(k)$ is the creation operator of Bloch mode with wave number $k$ in the sublattice A, and $\omega(k)$ is the dispersion relation (quasi energy) of the effective topological bath given by [Fig.5(b)]
\begin{equation}
\omega(k)=(\pi+k)/T \; \; \;\; {\rm (mod \; 2 \pi/T}).
\end{equation} 
The propagator $\hat{U}_2$, describing the evolution of the system in the third stage, from $t=2T_1$ to $t=T$, reads
\begin{equation}
\hat{U}_2=\exp(-i \hat{H}_{bs} T_2-i \hat{H}_0 T_2) 	\equiv \exp(-i \hat{H}_2 T)
\end{equation}
where $\hat{H}_0$ is the interaction Hamiltonian of discrete levels with the sites of sublattice A, i.e.
\begin{eqnarray}
\hat{H}_0 & = & \sum_{\alpha=1}^N \rho_{\alpha} \left\{ \hat{c}^{\dag}_{\alpha} \hat{A}_{n_{\alpha}}+H.c. \right\} \nonumber \\
& = &  \sum_{\alpha=1}^N \int dk \left\{ G_{\alpha}(k) \hat{c}^{\dag}_{\alpha} \hat{c}(k) +H.c. \right\}
\end{eqnarray}
with 
\begin{equation}
G_{\alpha}(k)= (\rho_{\alpha}/{\sqrt{2\pi}}) \exp(ikn_{\alpha}).
\end{equation}
 The propagator in one oscillation cycle, from $t=0$ to $t=T$, is given by
\begin{equation}
\hat{U}=\exp(-i  \hat{H}_2 T)  \exp(-i  \hat{H}_1 T) = \exp(-i \hat{H} T)
\end{equation}
where $\hat{H}$ is the effective Hamiltonian of the periodic (Floquet) dynamics. This means that, at times $t=0,T,2T,3T,...$, the time-periodic Floquet dynamics can be stroboscopically mapped  into the continuous dynamics of the effective time-independent Hamiltonian $\hat{H}$. Using the Baker-Campbell-Hausdorff formula and taking into account that, for a weak discrete-bulk coupling, the commutator $[ \hat{H}_1, \hat{H}_2]$ is negligible, the effective Hamiltonian takes the form given by Eq.(1) where $\omega(k)$ is defined by Eq.(23) [Fig.5(b)] and   
$g_{\alpha}(k)$ by Eq.(13) with $\kappa_{\alpha}=(T_2/T) \rho_{\alpha}$ (see Appendix C for technical details). An example of two-level decay into the anomalous Floquet topological insulator is shown in Fig.5(c). Note that, like for the previous example of the quantum Hall topological bath [Fig.4(d)], the decay of the initially-excited level $|1 \rangle$ is described by an almost exponential law, i.e. the decay is insensitive to the presence of the other discrete level $|2 \rangle$ owing to the unidirectional nature of Fano interference.
 
\section{Multi-particle quantum decay}

In previous sections we considered the decay dynamics of a single particle, so that the bosonic or fermionic nature of the particle does not affect the decay dynamics. However, in many particle systems the decay dynamics is influenced by the particle statistics, even for non-interacting particles \cite{r19,r20}. For example, fermions generally show a faster decay than bosons, as demonstrated in a recent experiment \cite{r20} where polarization-entangled photon states were used to emulate different particle statistics. In previous studies, the continuum was assumed to satisfy time reversal symmetry. A main question then arises: is the multi-particle decay dynamics affected by particle statistics when the decay occurs in a topological bath with broken time reversal symmetry?  Let us assume that $N$ indistinguishable particles are initially placed in the $N$ discrete levels $|1\rangle$, $|2 \rangle$, ... $|N\rangle$, and let us indicate by $P(t)$ the non-decaying probability, i.e. the probability that at time $t$ none of the $N$ particles have decayed into the topological continuum. As shown in \cite{r20}, the non-decaying probability $P(t)$ depends on the particle statistics and in the markovian approximation reads
\begin{equation}
P^{(ferm)}(t)= | {\rm det} (\mathcal{U}(t)) |^2
\end{equation}
for fermions, and 
\begin{equation}
P^{(bos)}(t)= | {\rm perm} (\mathcal{U}(t))|^2
\end{equation}
for bosons, where $\mathcal{U}(t)= \exp(-i \mathcal{H}t)$ is the propagator of the effective non-Hermitian Hamiltonian (6) and where $\rm {det} (\mathcal{U})$, $\rm {perm} (\mathcal{U})$ indicate the determinant and permanent of the matrix $\mathcal{U}$, respectively. If the decay occurs in a topological continuum sustaining unidirectional propagating modes, according to Eq.(14) the effective non-Hermitian Hamiltonian $\mathcal{H}$ is a lower triangular matrix, and thus also $\mathcal{U}(t)$ is a lower triangular matrix. In such a limiting case, the permanent and determinant of $\mathcal{U}$ do coincide, i.e. one has
\begin{equation}
P^{(ferm)}(t)=P^{(bos)}(t)=\exp(-\Delta t). 
\end{equation}
where we have set 
\begin{equation}
\Delta \equiv \sum_{\alpha=1}^N \frac{\kappa_{\alpha}^2 }{ v_{\alpha}}.
\end{equation}
Equation (30) indicates that, contrary to multiparticle quantum decay in a non-topological continuum, the non-decaying probability $P(t)$ is insensitive to particle statistics and always shows an exponential decay with a decay rate given by Eq.(31).

\section{Conclusion}
Fano interference among overlapping resonances is ubiquitous in the quantum mechanical decay process of two or more discrete states coupled to a common continuum. Fano interference is at the heart of important phenomena such as the existence of bound states in the continuum and fractional decay owing to destructive interference among different decay channels. In the many-body quantum decay process of non-interacting particles, Fano interference behaves different for bosonic and fermionic particles. Such previous results assume time reversal symmetry of the bath Hamiltonian, which is indeed the most common situation.
In this work we considered the process of multilevel quantum mechanical decay in a topological bath with broken time reversal symmetry sustaining unidirectional (chiral) propagating states, such as in quantum Hall or in Floquet topological insulators.  Such topological systems are nowadays available in different physical settings, such as in cold atoms or in photonic systems where synthetic gauge fields or Floquet dynamics can break time reversal symmetry \cite{r26,r32}.
 The main result of the present work is that the chiral nature of scattering states fully suppresses Fano interference among overlapping resonances, with a great impact into the
 multilevel quantum decay dynamics:  bound states in the continuum are suppressed, quantum decay is complete, and there are not signatures of particle statistics in the decay process. Nonetheless, some interesting features have been found, such as the appearance of high-order exceptional points, long quiescent dynamics followed by a sudden fast decay, and the possibility to observe damped non-Hermitian Bloch oscillations.  Our results disclose important novel physical behaviors in the quantum mechanical decay process when the underlying bath is a topological continuum and are expected to stimulate further theoretical and experimental research, bringing in close connection two apparently different areas of research: topological phases of matter and resonance physics.
 
\appendix

\section{Calculation of the effective non-Hermitian Hamiltonian in a continuum with unidirectional transport}
In this Appendix we derive the expression (14) of the non-Hermitian matrix elements $\mathcal{H}_{\alpha,\beta}$ given in the main text. For the sake of definiteness, we assume that $ (d \omega/ dk)$ is always positive in the entire Brillouin zone $-\pi \leq k < \pi$, corresponding to the topological bath sustaining forward-propagating modes solely [see Fig.1(d)], however a similar analysis holds for a topological continuum with  $ (d \omega/ d k)<0$. Substitution of Eq.(13) into Eq.(7) yields
\begin{eqnarray}
\Delta_{\alpha, \beta} & = & \frac{\kappa_{\alpha} \kappa_{\beta}}{2 \pi} \int_{- \pi}^{\pi} \left\{ dk \exp[ik (n_{\alpha}-n_{\beta})] \right. \times \nonumber \\
& \times & \left.  \int_0^{\infty} d \tau \exp [i \omega_{\beta} \tau -i \omega(k) \tau] \right\}
\end{eqnarray} 
Taking into account the identity
\begin{equation}
\int_0^{\infty} d \tau \exp(-i \Omega \tau)=\pi \delta (\Omega)-i \mathcal{P} \left(  \frac{1}{\Omega} \right)
\end{equation}
where $\mathcal{P}$ denotes the Cauchy principal value of the integral, from Eqs.(A1) and (A2) one obtains
\begin{eqnarray}
\Delta_{\alpha, \beta} & = & \frac{\kappa_{\alpha} \kappa_{\beta}}{2} \int_{- \pi}^{\pi}  dk \exp[ik (n_{\alpha}-n_{\beta})] \delta \left( \omega(k)-\omega_{\beta }\right)  + \nonumber \\
&-&   i \frac{\kappa_{\alpha} \kappa_{\beta}}{2 \pi} \mathcal{P} \int_{-\pi}^{\pi} dk \frac{\exp[ik (n_{\alpha}-n_{\beta})]}{\omega(k)-\omega_{\beta}}.
\end{eqnarray} 
Since the frequency $\omega_{\beta}$ of the discrete state $| \beta \rangle$ is embedded into the continuous spectrum of $\hat{H}_{bath}$ and $\omega(k)$ is an increasing function of $k$, there exists one and only one Bloch wave number $k_{\beta}$ such that $\omega(k_{\beta})=\omega_{\beta}$ [see Fig.1(d)]. Hence the first integral on the right hand side of Eq.(A3) reads
\begin{equation}
 \int_{- \pi}^{\pi}  dk \exp[ik (n_{\alpha}-n_{\beta})] \delta \left( \omega(k)-\omega_{\beta }\right) = \frac{\exp[ik_{\beta} (n_{\alpha}-n_{\beta})]}{v_{\beta}}  
\end{equation}
where we have set $v_{\beta} \equiv (d \omega / dk)_{k_{\beta}}$. Moreover, since we consider a broad and structureless continuum, in the second integral on the right hand side of Eq.(A3) we can assume $\omega(k) \simeq \omega_{\beta}+v_{\beta}(k-k_{\beta})$ for the dispersion relation; such as assumption is consistent with the weak-coupling and markovian approximations used to derive the effective non-Hermitian equations (5-7) given in the main text.  Under such an assumption, the principal value integral on the right hand side of Eq.(A3) reads 
\begin{equation}
\mathcal{P} \int_{-\pi}^{\pi} dk \frac{\exp[ik (n_{\alpha}-n_{\beta})]}{\omega(k)-\omega_{\beta}} \simeq \mathcal{P} \int_{-\pi}^{\pi} dk \frac{\exp[ik (n_{\alpha}-n_{\beta})]}{v_{\beta}(k-k_{\beta})}.
\end{equation}
For a broad continuum, corresponding to a large group velocity $v_{\beta}$, the integral on the right hand side in Eq.(A5) can approximately computed in complex plane using the residue theorem by extending  the integral from $-\infty$ to $\infty$. Taking into account that
\begin{equation}
\mathcal{P} \int_{-\infty}^{\infty} dx \frac{\exp(i \theta x)}{x-\xi}= \left\{ 
\begin{array}{cc}
0 & \theta=0 \\
i \pi \exp(i \theta \xi) & \theta >0 \\
-i \pi \exp(i \theta \xi) & \theta <0
\end{array}
\right.
\end{equation} 
one finally obtains
\begin{equation}
\Delta_{\alpha, \beta}= \left\{
\begin{array}{cc}
\frac{ \kappa_{\beta}^2}{2 v_{\beta}} & n_{\alpha}=n_{\beta} \\
\frac{\kappa_{\alpha} \kappa_{\beta}}{v_{\beta}} \exp[i k_{\beta}(n_{\alpha}-n_{\beta})] & n_{\alpha} > n_{\beta} \\
0 &  n_{\alpha} < n_{\beta}.
\end{array}
\right.
\end{equation}
Substitution of Eq.(A7) into Eq.(6) yields Eq.(14) given in the main text. 
\section{Effective non-Hermitian Hamiltonian in a tight-binding continuum with time reversal symmetry}
We consider the system shown in Fig.1(b), where $N$ discrete levels are side-coupled to a tight-binding continuum (a quantum wire) with time reversal symmetry and a dispersion relation $\omega=\omega(k)$ given by Eq.(9). The elements of the effective non-Hermitian Hamiltonian $\mathcal{H}$, defined by Eqs.(6) and (7), can be calculated in a  closed form and read explicitly (see \cite{r29} for more details)
\begin{equation} 
\mathcal{H}_{\alpha, \beta}=\omega_{\alpha} \delta_{\alpha,\beta}-i \kappa_{\alpha} \kappa_{\beta} i^{|n_{\alpha}-n_{\beta}|} \frac{\left( \sqrt{4 \kappa^2-\omega_{\beta}^2} +i \omega_{\beta}\right)^{|n_{\alpha}-n_{\beta}|}}{(2 \kappa)^{|n_{\alpha}-n_{\beta}|} \sqrt{4 \kappa^2 -\omega_{\beta}^2}}.
\end{equation}
In particular, for $\omega_{\alpha}=0$, i.e. when all discrete states have the same resonance frequency tuned at the center of the tight-binding lattice band, one simply obtains
\begin{equation} 
\mathcal{H}_{\alpha, \beta}=-i^{1+|n_{\alpha}-n_{\beta}|} \frac{\kappa_{\alpha} \kappa_{\beta}}{2 \kappa} .
\end{equation}
In case of two discrete levels ($N=2$) and for $\omega_1=\omega_2=0$, one bound state in the continuum, corresponding to a vanishing eigenvalue of $\mathcal{H}$, is found provided that $|n_{\alpha}-n_{\beta}|$ is an even number. Such a bound state shows some topological protection, i.e. it is robust against change of couplings $\kappa_{1,2}$ and disorder in the hopping rate of the quantum wire.
\section{Derivation of the effective Hamiltonian in the anomalous Floquet topological bath}
From Eq.(27) given in the main text and using the Baker-Campbell-Hausdorff formula, the effective Hamiltonian $\hat{H}$ reads
\begin{eqnarray}
\hat{H} & = & \hat{H}_1+\hat{H}_2+ \frac{iT}{2} \left[ \hat{H}_1, \hat{H}_2 \right]+ \\
& + & \frac{T^2}{12} \left(  \left[ \hat{H}_2, [\hat{H}_1,\hat{H}_2]  \right]- \left[ \hat{H}_1, [\hat{H}_1,\hat{H}_2]  \right] \right)+... \nonumber
\end{eqnarray}
where 
\begin{equation}
\hat{H}_1=\hat{H}_{bath}+\frac{2T_1}{T} \hat{H}_{bs} \;,  \;\;  \hat{H}_2=\frac{T_2}{T} (\hat{H}_{bs}+\hat{H}_0 )
\end{equation}
and where the dots denote terms involving higher-order commutators. For a weak discrete-continuum coupling, we show below that the commutator $ \left[ \hat{H}_1, \hat{H}_2 \right]$ is negligible, so that one has
\begin{equation}
\hat{H}  \simeq  \hat{H}_1+\hat{H}_2=\hat{H}_{bath}+\hat{H}_{bs}+\hat{H}_{int}
\end{equation}
where $\hat{H}_{bs}$ and $\hat{H}_{bath}$ are given by Eqs.(21) and (22), respectively, 
\begin{equation}
\hat{H}_{int}=\frac{T_2}{T} \hat{H}_0=\sum_{\alpha=1}^N \int dk \left\{ g_{\alpha}(k) \hat{c}^{\dag}_{\alpha} \hat{c}(k) + H.c. \right\}
\end{equation} 
and $g_{\alpha}(k)=(T_2/T) G_{\alpha}(k)$.\\ Let us calculate the commutator $ \left[ \hat{H}_1, \hat{H}_2 \right]$. We assume a small discrete-continuum coupling 
and an interaction time $T_2$, over one oscillation cycle, much smaller than the oscillation period $T$.  Under such an assumption and since $\hat{H}_{bath}$ and $\hat{H}_{bs}$ commute, one readily obtains
\begin{equation}
 \left[ \hat{H}_1, \hat{H}_2 \right] \simeq \frac{T_2}{T} \left[ \hat{H}_{bath}+\hat{H}_{bs}, \hat{H}_0 \right].
\end{equation}
For the sake of definiteness, let us assume bosonic commutation relations for the particle creation/annihilation operators, i.e. $[\hat{c}_{\alpha}, \hat{c}^{\dag}_{\beta}]= \delta_{\alpha, \beta}$, 
$[\hat{c}_{\alpha}, \hat{c}_{\beta}]=[\hat{c}^{\dag}_{\alpha}, \hat{c}^{\dag}_{\beta}]=0$, $[ \hat{c}(k), \hat{c}^{\dag}(k^{\prime}]=\delta(k-k^{\prime})$, and $[\hat{c}(k), \hat{c}^(k^{\prime}]=[ \hat{c}^{\dag}(k), \hat{c}^{\dag}(k^{\prime}]=0$. Using Eqs.(21), (22) and (25), the commutator on the right hand side of Eq.(C5) can be calculated after some straightforward algebra, yielding
\begin{eqnarray}
 \frac{iT}{2} \left[ \hat{H}_1, \hat{H}_2 \right] &  \simeq & \frac{T}{2} \sum_{\alpha=1}^{N} \int dk \left\{ i [ \omega_{\alpha}-\omega(k)] g_n(k) \hat{c}^{\dag}_{\alpha} \hat{c}(k)  \right. \nonumber \\
 & + & \left. H.c. \right\}
\end{eqnarray}
In the weak coupling approximation, energy conservation implies that the propagating states in the continuum excited in the decay process are those with Bloch wave number satisfying the condition $\omega(k) \simeq \omega_{\alpha}$; hence, if we set  $\omega(k) \simeq \omega_{\alpha}$ on the right hand side of Eq.(C6) under the sign of the integral, it follows that the correction introduced by the commutator  $(i T/2) \left[ \hat{H}_1, \hat{H}_2 \right]$ to $\hat{H}_{int}$ can be neglected.\\
Finally, let us notice that, if the discrete levels are degenerate in frequency, i.e. $\omega_{\alpha}$ is independent of $\alpha$, after a gauge transformation one can assume $\omega_{\alpha}=0$ and thus $\hat{H}_{bs}=0$. In this case, the condition $T_2 \ll T$ can be relaxed and the effective Hamiltonian $\hat{H}$, stroboscopically describing the Floquet dynamics, is again given by Eq.(C3).

\end{document}